\documentclass[english,aps]{revtex4}

\makeatletter

\providecommand{\tabularnewline}{\\}

\usepackage{babel}
\makeatother
\begin{document}

\title{Calculating the hydrogen molecule ion using the two particle Schr\"odinger
equation}

\author{Detlef Schmicker, Ludgeriplatz 31, 47057 Duisburg, Germany, 28th
Mart 2005}

\email{d.schmicker@physik.de}

\homepage{http://physik.de}

\begin{abstract}
We calculate the hydrogen molecule ion from the two particle Schr\"odinger
equation. Therefore a very simple two particle basis set is chosen.
We suggest this ansatz to be used to solve the \char`\"{}two electron
one phonon\char`\"{} three particle wave-function of a BCS superconductor.
Possibly it can give hints for high temperature superconductors.
\end{abstract}
\maketitle

\section{Introduction}

Quantum mechanical calculations of rotational, vibrational and electronic
excitations in molecules are usually done in the Born Oppenheimer
approximation \cite{Landau}. The electron-phonon system of crystals
is usually calculated in this approximation, too. Coupling of vibrations
and electronic excitations is calculated due to perturbation theory.
In this article we take the hydrogen molecule ion as a very simple
example of a system with rotational, vibrational and electronic excitations
and suggest a way of calculating it without using the Born Oppenheimer
approximation. Therefore we will use two particle wave-functions.
The same concept may be used to calculate electron-phonon systems
of crystals.

\section{Calculation}

We use the Schr\"odinger equation

\begin{equation}
H|\varphi>=e|\varphi>\label{1}\end{equation}

for the hydrogen molecule ion. The problem has three particles, but
it can be reduced to a two particle problem in a similar way, as the
hydrogen atom can be reduced from a two particle problem to a one
particle problem \cite{Landau}. The Hamiltonian has the form\begin{equation}
H=-\frac{\hbar}{2m_{a}}\Delta_{a}-\frac{\hbar}{2m_{b}}\Delta_{b}-\frac{\hbar}{2m_{c}}\Delta_{c}+\frac{e_{a}e_{b}}{|\overrightarrow{r_{a}}-\overrightarrow{r_{b}}|}+\frac{e_{b}e_{c}}{|\overrightarrow{r_{b}}-\overrightarrow{r_{c}}|}+\frac{e_{a}e_{c}}{|\overrightarrow{r_{a}}-\overrightarrow{r_{c}}|}\:.\label{2}\end{equation}

$\Delta_{a}$, $\Delta_{b}$ and $\Delta_{c}$ are the Laplace operators
with respect to the positions of the particles $\overrightarrow{r_{a}}$,
$\overrightarrow{r_{b}}$ and $\overrightarrow{r_{c}}$, respectively.
We introduce the center of mass\begin{equation}
\overrightarrow{R}=\frac{m_{a}\overrightarrow{r_{a}}+m_{b}\overrightarrow{r_{b}}+m_{c}\overrightarrow{r_{c}}}{m_{a}+m_{b}+m_{c}}\label{3}\end{equation}

and get new variables\begin{equation}
\overrightarrow{r_{a}}=\overrightarrow{r_{1}}+\overrightarrow{R}\,,\,\overrightarrow{r_{b}}=\overrightarrow{r_{2}}+\overrightarrow{R}\:,\label{4}\end{equation}
\begin{equation}
\overrightarrow{r_{c}}=-\frac{m_{a}\overrightarrow{r_{a}}+m_{b}\overrightarrow{r_{b}}}{m_{c}}=-\frac{m_{a}(\overrightarrow{r_{1}}+\overrightarrow{R})+m_{b}(\overrightarrow{r_{2}}+\overrightarrow{R})}{m_{c}}\:.\label{5}\end{equation}

We are only interested in solutions, which do not depend on the motion
of the center of mass, therefore we use $\overrightarrow{R}=0$\begin{equation}
\overrightarrow{r_{a}}=\overrightarrow{r_{1}}\,,\,\overrightarrow{r_{b}}=\overrightarrow{r_{2}}\,,\,\overrightarrow{r_{c}}=-\frac{m_{a}\overrightarrow{r_{1}}+m_{b}\overrightarrow{r_{2}}}{m_{c}}\:.\label{6}\end{equation}

Let $m_{a}=m_{e}$ be the electron mass, $m_{b}=m_{c}=m_{p}$ be the
proton mass, $e_{a}=-e$ the electron charge and $e_{b}=e_{c}=e$
the proton charge. With $m_{a}\ll m_{c}$ and $m_{b}=m_{c}$ we get\begin{equation}
\overrightarrow{r_{c}}=-\overrightarrow{r_{2}}\:.\label{7}\end{equation}

This is not a Born Oppenheimer approximation, it is just a simplification
of the calculation. The value of $r_{c}$ can be wrong by 0.0005 of
usual distances in the hydrogen molecule ion. Within our calculation
of the matrix elements this will be a small error with respect to
the integration errors from Monte-Carlo integration.

Using the new variables $\overrightarrow{r_{1}}$ and $\overrightarrow{r_{2}}$
we get a two particle Hamiltonian\begin{eqnarray}
H & = & -\frac{\hbar}{2m_{e}}\Delta_{1}-\frac{\hbar}{4m_{p}}\Delta_{2}+\frac{e^{2}}{2|\overrightarrow{r_{2}}|}-\frac{e^{2}}{|\overrightarrow{r_{1}}+\overrightarrow{r_{2}}|}-\frac{e^{2}}{|\overrightarrow{r_{1}}-\overrightarrow{r_{2}}|}\:.\label{8}\end{eqnarray}

$\Delta_{1}$, $\Delta_{2}$ are the Laplace operators with respect
to $\overrightarrow{r_{1}}$ and $\overrightarrow{r_{2}}$, respectively.
Due to constrains the mass moved by changing $\overrightarrow{r_{2}}$
is $2m_{p}$. With this Hamiltonian we calculate the matrix elements
for the Schr\"odinger equation for a chosen two particle basis-function
set. We use the linear combination ansatz and the variation theorem\begin{equation}
\varphi(\overrightarrow{r_{1}},\overrightarrow{r_{2}})=\Sigma_{i}c_{i}\varphi_{i}(\overrightarrow{r_{1}},\overrightarrow{r_{2}})\:.\label{9}\end{equation}

From the expectation value $e$ of the energy\begin{equation}
<\varphi|H|\varphi>=e<\varphi|\varphi>\:,\end{equation}

we get \begin{equation}
\int dV_{1}dV_{2}\Sigma_{ij}c_{j}\varphi_{j}^{*}(\overrightarrow{r_{1}},\overrightarrow{r_{2}})Hc_{i}\varphi_{i}(\overrightarrow{r_{1}},\overrightarrow{r_{2}})=\int dV_{1}dV_{2}\Sigma_{ij}c_{j}\varphi_{j}^{*}(\overrightarrow{r_{1}},\overrightarrow{r_{2}})\, e\, c_{i}\varphi_{i}(\overrightarrow{r_{1}},\overrightarrow{r_{2}})\:,\end{equation}

and write it as

\begin{equation}
\Sigma_{ij}c_{j}H_{ij}c_{i}=e\Sigma_{ij}c_{j}S_{ij}c_{i}\label{eq:erwartungswert}\end{equation}

with\begin{equation}
H_{ij}=\int dV_{1}dV_{2}\varphi_{i}H\varphi_{j}\;,\; S_{ij}=\int dV_{1}dV_{2}\varphi_{i}\varphi_{j}\:.\label{HamiltonOverlap}\end{equation}
The integrals must be calculated over two 3D-volumes, as $\varphi_{i}$
are two particle basis-functions. From the variation theorem the energy
$e$ is a minimum with respect to all parameters $c_{j}$

\begin{equation}
\frac{\partial}{\partial c_{j}}\:\frac{\Sigma_{ij}c_{j}H_{ij}c_{i}}{\Sigma_{ij}c_{j}S_{ij}c_{i}}=0\:.\label{eq:partdiff}\end{equation}
With this ansatz the Schr\"odinger equation looks like\begin{equation}
\Sigma_{i}H_{ij}c_{i}=e\;\Sigma_{i}S_{ij}c_{i}\:.\label{matrixschroedinger}\end{equation}

This is a general eigenvalue problem which can be solved numerically.
Also the integrals $H_{ij}$ and $S_{ij}$ can be solved numerically,
for example by Monte Carlo integration.

\section{Basis-functions for solving the two particle Schr\"odinger equation}

The choice of the two particle basis-functions $\varphi_{i}(\overrightarrow{r_{1}},\overrightarrow{r_{2}})$
will be evaluated now. If two particle electron wave-functions are
needed, often the ansatz $\varphi_{pq}(\overrightarrow{r_{1}},\overrightarrow{r_{2}})=\psi'_{p}(\overrightarrow{r_{1}})\phi'_{q}(\overrightarrow{r_{2}})$
is used. In our case this ansatz will be very bad, because one particle
is a (quasi) proton, the other an electron. The motion of the electron
is not independent from the motion of the protons, as the electrons
will have highest probability density near the protons. A simple approach
could be\begin{equation}
\varphi_{pq}(\overrightarrow{r_{1}},\overrightarrow{r_{2}})=k_{pq}\psi_{p}(\overrightarrow{r_{1}})\left(\phi_{q}(\overrightarrow{r_{2}}-\overrightarrow{r_{1}})+s\phi_{q}(\overrightarrow{r_{2}}+\overrightarrow{r_{1}})\right)\:.\label{eq:seperation}\end{equation}

We will use $\psi_{p}$ and $\phi_{q}$ as radial symmetric functions.
The combination of the indices $p$ and $q$ correspond to the index
$i\rightarrow pq$ in (\ref{matrixschroedinger}). $k_{pq}$ is the
normalization constant and will be calculated from the overlap integrals
$\int dV_{1}dV_{2}\varphi_{pq}\varphi_{pq}=1$. $\left(\phi_{q}(\overrightarrow{r_{2}}-\overrightarrow{r_{1}})+s\phi_{q}(\overrightarrow{r_{2}}+\overrightarrow{r_{1}})\right)$
describes the electron part of the basis-functions. The highest probability
density of the electron is near both protons. $s$ is $+1$ or $-1$
characterizing the bonding and anti-bonding wave-functions of the
electron \cite{Landau}. We will only use $s=1$ for bonding wave-functions
and use simple basis-functions for $\phi_{q}$ and $\psi_{p}$\begin{eqnarray}
\psi_{p}(\overrightarrow{r}) & = & f_{p}(\overrightarrow{r})e^{-w_{proton}(|\overrightarrow{r}|-r_{0})^{2}}\:,\nonumber \\
\phi_{0}(\overrightarrow{r}) & = & e^{-w_{electron}(|\overrightarrow{r}|)^{2}}\:.\label{eq:basis functions}\end{eqnarray}

$r_{0}$ is the half of the proton-proton distance. As functions $f_{p}$
we use:\[
f_{0}(\overrightarrow{r})=1\quad,\quad f_{1}(\overrightarrow{r})=\frac{x}{|\overrightarrow{r}|}\quad,\quad f_{2}(\overrightarrow{r})=(|\overrightarrow{r}|-r_{0})\quad,\quad f_{3}(\overrightarrow{r})=f_{2}(\overrightarrow{r})f_{1}(\overrightarrow{r})\quad,\]
\begin{equation}
with\;\overrightarrow{r}=\left(\begin{array}{c}
x\\
y\\
z\end{array}\right)\:.\end{equation}

$\psi_{0}$ should be similar to the wave-function for $l=0\,,\, n=0$
(rotational and vibrational ground stage), $\psi_{1}$ is similar
to $l=1\,,\, n=0$, $\psi_{2}$ is similar to $l=0\,,\, n=1$, $\psi_{3}$
is similar to $l=1\,,\, n=1$. $\psi_{0}$ is constructed as ground
wave-function of a harmonic oscillator, $\psi_{2}$ is the first excitation
of a harmonic oscillator with respect to the radial dependency ($r_{0}$
is the mean distance of the protons from the center of mass) \cite{Landau}.
$f_{1}(\overrightarrow{r})$ describes the spheric harmonics for $l=1$.
$\phi_{0}(\overrightarrow{r})$ is a Gaussian type ground wave-function
for the electron. With this basis-functions we choose approximate
parameters in atomic units\begin{eqnarray}
w_{proton} & = & 18\:,\nonumber \\
w_{electron} & = & 0.45\:,\label{eq:parameters}\\
r_{0} & = & 1.0\:.\nonumber \end{eqnarray}

This parameters can be calculated by searching the minimum of the
ground state energy from equation (\ref{matrixschroedinger}) with
respect to $w_{proton}$, $w_{electron}$ and $r_{0}$. This is not
done within this article, as the computational power was not available.
We only did some test calculations to check, if we are near the minimum.
The parameters are taken from known properties of the hydrogen molecule
ion \cite{ExpRes}. 

\begin{table}
\begin{center}\begin{tabular}{|c|c|c|c|c|}
\hline 
Eigenvalue {[}a.u.{]}&
$c_{00}$&
$c_{10}$&
$c_{20}$&
$c_{30}$\tabularnewline
\hline
\hline 
-0.521112&
-0.99955&
0&
-0.0299897&
0\tabularnewline
\hline 
-0.520852 &
0&
-0.999777&
0&
-0.0211105\tabularnewline
\hline 
-0.505818&
-0.198128&
0&
0.980176&
0\tabularnewline
\hline 
 -0.505597&
0&
0.190136&
0&
-0.981758\tabularnewline
\hline
\end{tabular}\end{center}

\caption{\label{cap:Results-of-sample}Results of sample calculation}
\end{table}
 The results for the four eigenvalues and eigenvectors are shown in
table \ref{cap:Results-of-sample}. They are not very good compared
to other calculations, but of cause the choice of the basis set is
very simple. The ground state wave-function is close to $\varphi_{00}$,
the first excitation is the rotational excitation with a wave-function
close to $\varphi_{10}$. The second excitation is the vibrational
excitation with a wave-function close to $\varphi_{20}$. The calculated
energy for the first rotational excitation is $0.00026\, a.u.=0.007\, eV$.
The calculated energy for the first vibrational excitation is $0.0153\, a.u.=0.41\, eV$.
The values can be compared to literature \cite{ExpRes}. The energy
of the vibrational excitation is calculated to high. The literature
value is about $0.27\, eV$. One can see from table \ref{cap:Results-of-sample},
that the eigenfunctions are not very good. In the linear combination
also $c_{00}$ differs from zero.

\section{Discussion}

The suggested basis-functions (\ref{eq:seperation}) can be used to
calculate the two particle problem of the hydrogen molecule ion. As
result we get two particle wave-functions for the ground state, the
first rotational excitation and the first vibrational excitation.

The chosen basis-functions are very simple functions and the basis
set is very small. They can of cause be improved in a number of ways.
Especially for the electron part of the wave-function we did not even
include basis-functions for electronic excitations. 

Additionally the separation (equation: \ref{eq:seperation}) does
not allow electron phonon interaction, as the electron wave-function
does not depend on the protons. One will need electron basis-functions,
which include the proton-proton distance $2|\overrightarrow{r_{2}}|$
to include electron-phonon interaction. This is not a principle problem,
as the separation is only used to construct simple basis-functions.
The separation is not needed for the calculation in any way. With
enough computational power it should be possible to use a subset of
a complete set of basis-functions. With the use of such a subset,
no artificial approximations are needed to solve the Schr\"odinger equation
and the precision can be increase by increasing the number of basis-functions
in the subset.

\section{Outlook}

This ansatz can be transferred to crystal lattices. The aim would
be to calculate the electron-phonon system in a one electron, one
phonon (more general: vibrational excitation) approximation ab initio.
In a next step a ''two electron one phonon'' approximation could
be used to calculate the electron-phonon system of a BCS superconductor.
With the choice of a subset of a complete set of three particle basis-functions
no additional approximations will be needed (but the limited number
of basis-functions in the subset). It would be quite interesting to
see, how this wave-functions look like. 

Possibly this ansatz can give new ideas, how to describe high temperature
superconductors.

\section{Resources}

The source code of this calculation is available \cite{SourceCode}.
Please contact the author in case you want to work on the topic.

\end{document}